# Enhanced IoV Security Network by Using Blockchain Governance Game*

SONG-KYOO KIM

**ABSTRACT**

This paper deals with the design of the secure network in an Enhanced Internet of Vehicles by using the Blockchain Governance Game (BGG). The BGG is a system model of a stochastic game to find best strategies towards preparation of preventing a network malfunction by an attacker and the paper applies this game model into the connected vehicle security. Analytically tractable results for decision-making parameters enable to predict the moment for safety operations and to deliver the optimal combination of the number of reserved nodes with the acceptance probability of backup nodes to protect a connected car. This research helps for whom considers the enhanced secure IoV architecture with the BGG within a decentralized network.

**Keywords:** IoT security; Internet of Vehicles; IoV; connected car, Blockchain Governance Game; mixed game; stochastic model; fluctuation theory; 51 percent attack

## I. INTRODUCTION

The Enhanced BIoV (EBIoV) network [1] is emphasized in this paper. The EBIoV is the IoV network architecture based on the Edge Computing [2-3] with enabling the Blockchain Governance Game (BGG) [4] for improving the network security [1]. Atypical BIoV could be applied to trace the provenance of spare parts back through every step in the supply chain to its original manufacture date and location [1, 5]. Car manufacturers concern that service centers and garages are knowingly fitting counterfeit spare parts to their vehicles of customers [2]. Identifying geniality of car parts by using the EBIoV has been introduced in the previous research [6] and it suggests that the network within car components is considered as the Edge network and the monitoring equipment in car service centers are Fog level and the database in the headquarter (HQ) is the level of the cloud network (see Fig.1) [2-3, 7].

---

* This paper is the abridged summary of the accepted paper.

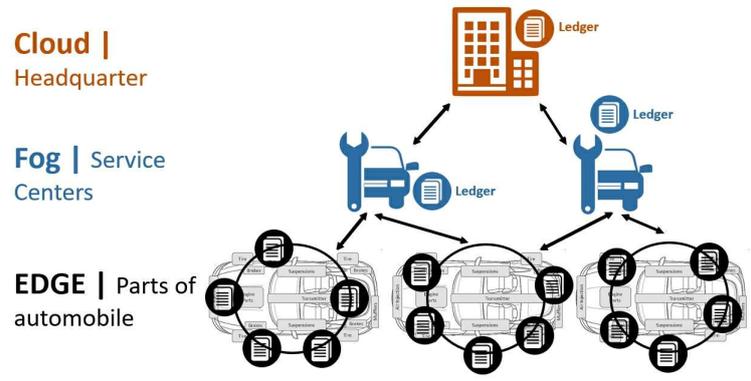

**Fig 1.** Adopting Fog Computing in Auto Services in Blockchain [1]

This paper provides the mathematical functional of the EBIoV network architecture for enhancing the securities from attackers. The EBIoV security regarding counterfeits of the car parts already has been studied [1] and this research is focused on avoiding atypical IoV attack in a decentralized network (i.e., the 51 percent attack). In the BGG [4], an attacker and a defender compete each other by building blocks in private and public chains as the sequence of stages to generate ledgers. The historical strategies and the probabilistic stage transitions can be observed by both an attacker and a defender. Hence, the interaction between the attacker and the defender can be modeled as a stochastic game [4, 8].

## II. STOCHASTIC GAME FOR BIoV NETWORK SECURITY

The Blockchain Governance Game (BGG) [4] has been tried to be adapted the BIoV network architecture to defend against the attacks. This system model consists of one attacker (i.e., the miner which intends to fork a private chain) and one defender (the miner which honestly mines on the public chain) [8]. This explicit function (Theorem BGG-1) from the BGG (Blockchain Governance Game) gives the predicted moment one step before the 51 percent attack [4]. The proposed BIoV network structure is considered [1] and the components in a vehicle, the equipment of a service center and the HQ database are hooked up as one Blockchain (see Fig. 2).

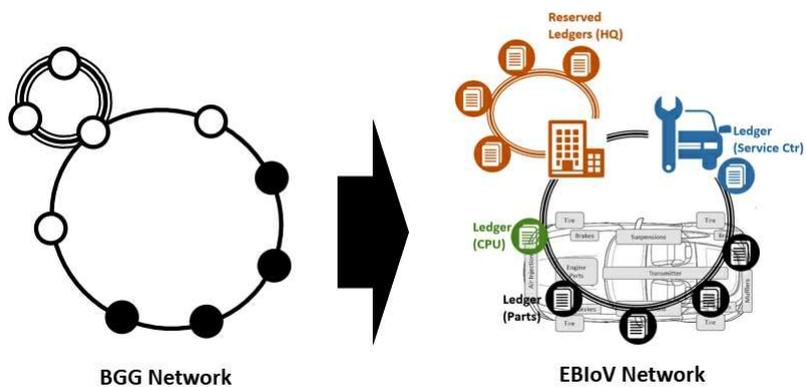

**Fig 2.** Adapting BGG for the EBIoV architecture [1]

Each car component beside the CPU (smart controller) generates the value based on its mechanical action and the generated values are sharing with all other components including assigned service centers and a company headquarter. Each number from a car component is unique and randomly generated. Unlike conventional Blockchain networks, the EBIoV network does not have any reward system which requires a heavy computational power for being a miner to generate ledgers. The Verifiable Random Function (VRF) is applied to select the node for generating ledgers [9] and it is a cryptographic primitive that maps inputs to verifiable pseudorandom outputs [10-11]. One Blockchain company has pioneered for using the VRF to perform secret cryptographic solution to select committees to run the consensus protocol [9]. By applying the VRF, all nodes will have an equal chance to become a miner for generating ledgers without requiring a heavy computational power. The mechanism for protecting the Blockchain is exactly same as the BGG.

To apply the BGG into the BIoV network structure, the antagonistic game of two players (called "A" and "H") are introduced to describe the Blockchain network in a connected car as a defender and an attacker. Both players compete to build the blocks either for honest or false ones. Let $(\Omega, \mathcal{F}(\Omega), P)$ be probability space $\mathcal{F}_A$, $\mathcal{F}_H$, $\mathcal{F}_\tau \subseteq \mathcal{F}(\Omega)$ be independent $\sigma$-subalgebras. This game is a stochastic process $\mathcal{A}_\tau \otimes \mathcal{H}_\tau$ describing the evolution of a conflict between players A and H known to an observation process $\mathcal{T} = \{\tau_0, \tau_1, \dots\}$. The game is ended when on the $k$-th observation epoch $\tau_k$, the collateral building blocks to player H (or A) exceeds more than the half of the total nodes $M$ in the regular operation or player A exceeds more than $\left(\frac{M}{2}\right) + B$ nodes under the safety mode. To further formalize the game, the *exit index* is introduced:

$$\nu := \inf\left\{k : A_k = A_0 + X_1 + \cdots + X_k \geq \left(\tfrac{M}{2}\right) + B\right\}, \quad (1)$$

$$\mu := \inf\left\{j : H_j = H_0 + Y_1 + \cdots + Y_j \geq \left(\tfrac{M}{2}\right)\right\}. \quad (2)$$

where $B$ is the number of the reserved honest nodes from a headquarter (HQ) which is depends on the availability from the HQ. The joint functional of the BIoV network model is as follows:

$$\Phi\left(\zeta; \left\lceil\tfrac{M}{2}\right\rceil + B, \left\lceil\tfrac{M}{2}\right\rceil\right) \quad (3)$$
$$= \mathbb{E}\left[\zeta^\nu \cdot g_0^{A_{\nu-1}} \cdot g_1^{A_\nu} \cdot z_0^{H_{\nu-1}} \cdot z_1^{H_\nu} \mathbf{1}_{\{\nu < \mu\}} \Big| B\right]\Big|_{(g_0, g_1, z_0, z_1) \to 1}$$

where $M$ indicates the total number of nodes (or ledgers) in the BIoV network for each car (see Fig. 2).

## III. BLOCKCHAIN GOVERNANCE GAME STRATEGY & OPTIMIZATION PRACTICE

Let us consider a two-person mixed strategy game, and player H (i.e., a car company) is the person who has two strategies at the observation moment, one step before attackers complete to generate alternative chains with dishonest transactions. Player H has the following strategies:

. Players: $$\mathcal{N} = \{A, H\}, \tag{4}$$
. Strategy sets:
$$s_a = \{"NotBurst", "Burst"\},$$
$$s_h = \{"Regular", "Safety"\}.$$

Based on the above conditions, the general cost matrix at the prior time to be burst $\tau_{\nu-1}$ could be composed as follows:

**Table 1.** Cost matrix

|         | NotBurst $(1-q(s_h))$ | Burst $(q(s_h))$ |
|---------|------------------------|-------------------|
| Regular | 0                      | $V$               |
| Safety  | $c_b$                  | $c_b + V$         |

where $q(s_h)$ is the probability of bursting blockchain network (i.e., an attacker wins the game) and it depends on the strategic decision of player H:

$$q(s_h) = \begin{cases} \mathbb{E}\left[\mathbf{1}_{\{A_\nu \geq \frac{M}{2}\}}\right], & s_h = \{Regular\}, \\ \mathbb{E}\left[\mathbb{E}\left[\mathbf{1}_{\{A_\nu \geq \frac{M}{2}+B\}}\Big|B\right]\right], & s_h = \{Safety\}. \end{cases} \tag{5}$$

It is noted that the cost for the reserved nodes (i.e., the cost of "*Safety*" operation strategy by player H) should be smaller than the other strategy. The total number of reserved nodes $n$ that a company owns depends on the cost function and the optimal number of reserved nodes $n^*$ which supported by the HQ. Similarly, the optimal value $\rho^*$ is the success probability when the reserved honest nodes are supported from the HQ. The best combination $(n^*, \rho^*)$ could be found as follows:

$$(n^*, \rho^*) = \inf\{(n, \rho) \geq 0 : \mathfrak{S}_{\text{Reg}}(q^0) \geq \mathfrak{S}_{\text{Safe}}(n, \rho)\}. \tag{6}$$

The governance in the Blockchain is followed by the decision making parameter and it also means no action until the time $\tau_{\nu-1}$. An attacker still has the chance that all nodes are governed by an attacker if the attacker catches more than the half of nodes at $\tau_{\nu-1}$ (i.e., $\{A_{\nu-1} \geq \frac{M}{2}\}$). If the attacker catches the less than half of all nodes at $\tau_{\nu-1}$ (i.e., $\{A_{\nu-1} < \frac{M}{2}\}$), then the defender could run the safety mode to avoid the attack at $\tau_\nu$. The total cost for developing the enhanced BIoV network is as follows:

$$\mathfrak{S}(q^0; n, \rho)_{\text{Total}} = \mathbb{E}\left[\mathfrak{S}_{\text{Safe}}(n, \rho) \cdot \mathbf{1}_{\{A_{\nu-1} < \frac{M}{2}\}} + \mathfrak{S}_{\text{Reg}}(q^0) \cdot \mathbf{1}_{\{A_{\nu-1} \geq \frac{M}{2}\}}\right] \tag{7}$$

$$= \left\{c_{(n,\rho)}\left(1 - q^1_{(n,\rho)}\right) + \left(c_{(n,\rho)} + n\rho\right)q^1_{(n,\rho)}\right\}p_{A_{-1}} + B \cdot q^0(1 - p_{A_{-1}})$$

where

$$p_{A_{-1}} = \boldsymbol{P}\left\{A_{\nu-1} < \frac{M}{2}\right\} = \sum_{k=0}^{\lfloor \frac{M}{2} \rfloor} \boldsymbol{P}\{A_{\nu-1} = k\}. \tag{8}$$

Recall from (5), the probability of bursting Blockchain network (i.e., an attacker wins the game) under the memoryless properties becomes the Poisson compound process:

$$q(s_h) = \begin{cases} \sum_{k > \frac{M}{2}} \mathbb{E}\left[\mathbf{1}_{\{A_\nu = k\}}\right], & s_h = \{Regular\}, \\ \mathbb{E}\left[\mathbb{E}\left[\sum_{k > \frac{M}{2}+B} \mathbb{E}\left[\mathbf{1}_{\{A_\nu = k\}}\right] \bigg| B\right]\right], & s_h = \{Safety\}, \end{cases} \tag{9}$$

where

$$\mathbb{E}\left[\mathbf{1}_{\{A_\nu = k\}}\right] = \mathbb{E}\left[\mathbb{E}\left[\frac{\lambda_a \tau_\nu}{k!} \cdot e^{-\lambda_a \tau_\nu} \bigg| \tau_\nu\right]\right]. \tag{10}$$

A network security in a BIoV network for each car is considered in this subsection. The strategy for protecting the EBIoV is for supporting the additional nodes to give the less chance that an attacker catches blocks with false control requests. Since, the model of the BGG in the EBIoV network has been analytically solved, the values for the cost function and the calculations of the probability distributions are straight forward. However, it still requires the software implementation by using a programing language. Based on the above conditions, the LP (Linear Programing) model could be described as follows from (3.7) and (3.4):

Objective
minimizing $G = \mathfrak{S}(n, \rho)_{\text{Total}}$ (11)

Subject to
$n \geq \frac{c_{(n,\rho)}}{V \cdot q^0 - c_{(n,\rho)}};$ (12)

From (7), the total cost $\mathfrak{S}(n, \rho)_{\text{Total}}$ is as follows:

(13)
$$\mathfrak{S}(n, \rho)_{\text{Total}} = \left(c_{(n,\rho)}(1 - q_\eta^1) + (c_{(n,\rho)} + V)q_{(n,\rho)}^1\right)p_{A_{-1}}$$
$$+ V \cdot q^0(1 - p_{A_{-1}}).$$

The total cost $\mathfrak{S}(n, \rho)_{\text{Total}}$ could be minimized by given $(n, \rho)$ and the parameter set $(n^*, \rho^*)$ is the optimal combination of the reserved nodes which are supported by the HQ. The below illustration in Fig. 3 is atypical graph of an optimal result by using the BGG based BIoV (EBIoV) network based on the given conditions in Table I.

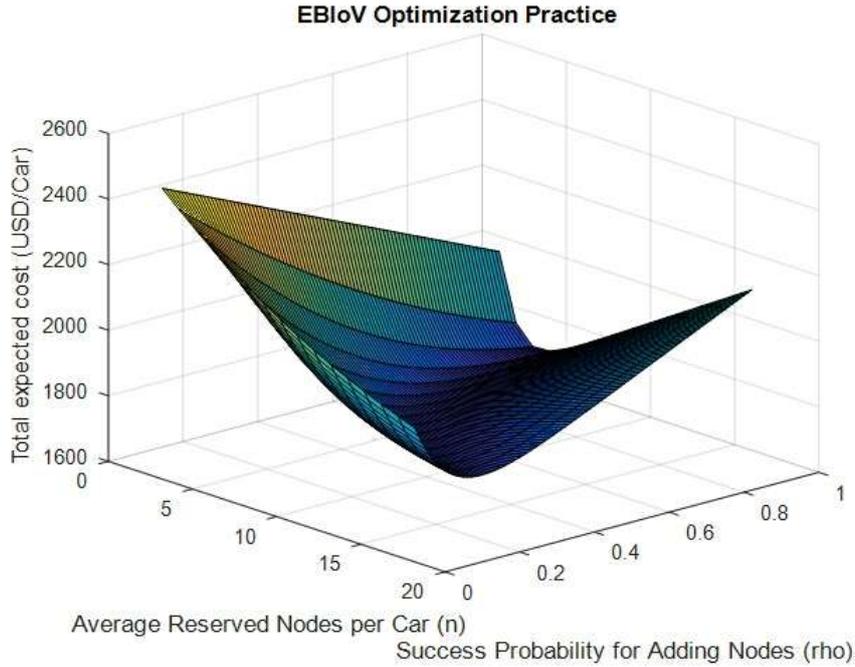
**Fig. 3.** Optimization Example for the EBIoV

The moment of requesting the additional nodes will be the time $\tau_{\nu-1}$ when is one step prior than the time when an attacker catches more than half of the whole blocks (i.e., $\tau_\nu$).

# V. CONCLUSION

This paper establishes the enhanced Blockchain based IoV network architecture by bringing a theoretical model in stochastic modeling. The Enhanced Blockchain enabled Internet of Vehicles (EBIoV) is an advanced secure IoT network architecture for protecting a connected car from attackers. This new architecture has been designed for a decentralized network by adapting the Blockchain Governance Game (BGG) to improve the connected car security. The BGG is a mathematically proven game model to develop optimal defense strategies to protect systems from attackers. The practical case in the paper demonstrates how an EBIoV network could be implemented for connected car securities. The EBIoV network is the first research that applies a BGG model into the IoV security domain. The BGG model shall be extended to various Blockchain based cybersecurity areas including IoT security and secured decentralized service network design.